\documentstyle[aps,epsfig]{revtex}
\def\be{\begin{equation}}
\def\ee{\end{equation}}
\begin{document}
 \title{On the entanglement entropy for a XY spin chain}
 \author{
Ingo Peschel\\
{\small Fachbereich Physik, Freie Universit\"at Berlin,} \\
{\small Arnimallee 14, D-14195 Berlin, Germany}}
 \maketitle
 \begin{abstract}
 The entanglement entropy for the ground state of a XY spin chain
 is related to the corner transfer matrices of the triangular Ising
 lattice and expressed in closed form.

\end{abstract}

\vspace{2cm}

In the study of the properties of quantum states the
entanglement entropy plays an important r$\hat{o}$le. It is
calculated from the reduced density matrix for part of the system 
and measures how many states are
connected across the interface between both parts. For
critical one-dimensional systems, it diverges logarithmically 
with the size of the subsystem, and the prefactor of the logarithm
involves the conformal anomaly c. \cite{Holzhey94,Calabrese04}. 
This has been checked numerically
for a number of solvable spin chains recently \cite{Vidal03,Latorre03}.
Analytical calculations exist for for the (critical) XX chain (equivalent to
free fermions hopping on a lattice) \cite{Jin03} and for the (non-critical) XY 
chain in a field \cite{Its04}, as well as for the ferromagnetic Heisenberg
chain \cite{Popkov04}. In the first two cases, the limit of a large subsystem 
was treated by using asymptotic properties of the Toeplitz matrices which
determine the density matrix eigenvalues. For the XY chain, the entropy
was thereby obtained as an integral over theta functions. In the present 
note we want to point out that the XY chain can also be treated by 
connecting it to a triangular Ising model and its corner transfer matrices.
Moreover, a much simpler formula for the entropy can be given (at least 
for part of the parameter space) which involves only elliptic integrals.\\

Our aim is to study the ground state $|0>$ of the XY Hamiltonian 
 \be
   H = -  \sum_{n} \;[\;(1+\gamma)\sigma_n^x \sigma_{n+1}^x +
              (1-\gamma) \sigma_n^y \sigma_{n+1}^y + 2h \sigma_n^z]
   \label{eqn:XY}
  \ee
 where the $\sigma_n^{\alpha}$ are Pauli matrices and $0 \leq \gamma
 \leq 1, h \geq 0$. This operator can be obtained from
  \be
  \hat{H} = - \sum_{n} \;[\;(1+\gamma)\tau_n^x -
              (1-\gamma) \tau_{n-1}^z \tau_n^x\tau_{n+1}^z + 
              2h \tau_n^z\tau_{n+1}^z]
  \label{eqn:XYD}
 \ee
 where the $\tau_n^{\alpha}$ are also Pauli matrices
 by the dual transformation $\tau_n^x=\sigma_{n-1}^x \sigma_n^x,
 \;\;\tau_n^z\tau_{n+1}^z=\sigma_n^z$. The ground state of $\hat{H}$
 will be denoted by $|\hat{0}>$. It is important to make this distinction
 because the dual transformation affects the entanglement properties.
 The bond variable $\sigma^z$ only registers the relative orientation
 of the neighbouring $\tau^z$-spins.\\
 
  Now it is known that $\hat{H}$ commutes with the symmetrized transfer 
 matrix $T$ of a triangular Ising model shown in Fig. 1
 \cite{Stephen72,Krinsky72,PeschelEmery81}. Moreover, $|\hat{0}>$ corresponds to the 
 maximal eigenvalue of $T$. The relation of the three
 coupling constants $K_i= \beta J_i$ of the triangular
 lattice and the chain parameters is 
 \be
   \gamma = 1/C_3, \;\; h= (S_1S_2C_3+C_1C_2S_3)/C_3
   \label{eqn:parameter}
  \ee
  where $S_i=sinh\;2 K_i$ and $C_i=cosh\;2 K_i$. The ordered region of
 the Ising system, $T < T_c$, corresponds to $h > 1$, the disordered
 one to $h < 1$.
 For $K_3 = 0$, one is dealing with a square lattice. Then $\gamma = 1$ 
 and $\hat{H}$ becomes the Hamiltonian of an Ising chain in a transverse
 field. This system was studied in \cite{Pescheletal99} by noting that the
 reduced density matrix $\hat{\rho}$ for a half-chain is the partition function 
 of a long Ising strip with a perpendicular cut. If the chain length resp. the
 strip width are much larger than the correlation length, this partition function 
 can be expressed in terms of infinite corner transfer matrices (CTM's).\\
 
 The same can be done for the triangular lattice. In this case one multiplies
 alternatingly CTM's with a central spin ($A_i$) and without one ($H_i$) as 
 shown in Fig. 2. These quantities were studied in \cite{Tsang77,Baxter93}. 
 The result is that $\hat{\rho}$ has the diagonal form
  \be
   \hat{\rho}= {\cal K} \exp{(- \sum_{j \geq 0}
 \hat{\varepsilon}_j c_j^{\dagger} c_j)}
   \label{eqn:rho1}
  \ee
with fermionic operators  $c_j^{\dagger}, c_j$ and single particle eigenvalues
  \be
  \hat{\varepsilon}_j = \left \{ \begin {array} {c} (2j+1)\; \varepsilon ,\;\;\; 
                                     T > T_c \; \;\;  \\[0.3cm]
                           2j\;\varepsilon ,\;\;\; T < T_c \; \;\; \end{array} \right.
  \label{eqn:epsilon1}
  \ee
 The factor $\varepsilon$ which sets the scale is $\varepsilon= \pi\;I(k')/I(k)$ where 
 $I(k)$ denotes the complete elliptic integral of the first kind and $k'=\sqrt{1-k^2}$. 
 The modulus $k$, finally, is the Onsager parameter for the triangular lattice and given 
 by eqn. (6.4.16) in Ref. \cite{Baxter82}. In terms of the chain parameters one finds, 
 using (\ref{eqn:parameter}) and choosing $K_1 =K_2$, that
 \be
  k= \left \{ \begin {array} {c} \sqrt{h^2+\gamma^2-1}\; /\; \gamma , \;\;\;h<1 \\ [0.3cm]
       \gamma\; / \;\sqrt{h^2+\gamma^2-1} ,\;\;\; h>1 \end{array} \right.
  \label{eqn:k}
 \ee
 To obtain the reduced density matrix $\rho$ for the XY chain, one notes that the
 dual transformation interchanges order and disorder. Therefore one has to 
 interchange the expressions for the eigenvalues in (\ref{eqn:epsilon1}) to obtain
 \be
   \varepsilon_j =\left \{ \begin {array} {c}2j\;\varepsilon ,\;\;\; h < 1\\[0.3cm]
                          (2j+1)\; \varepsilon ,\;\;\; h > 1 \end{array} \right.
    \label{eqn:epsilon2}
 \ee
 In the special case of the transverse Ising chain ($\gamma = 1$) this spectrum 
 coincides with the one obtained directly from (\ref{eqn:XYD}) and  (\ref{eqn:rho1}).
 The eigenvalue $\varepsilon_0 = 0$ describes the (asymptotic) degeneracy of the
 density matrix eigenfunctions for $h < 1$ connected with the spin inversion 
 $\sigma^x \rightarrow -\sigma^x$. \\
 
 This is valid for purely ferromagnetic interactions as well as for antiferromagnetic
 competing interactions in the Ising model. If we confine ourselves to the first case,
 $h$ and $\gamma$ satisfy $h^2 + \gamma^2 > 1$ and $k$ is real, $0 \leq k\leq 1$.
 Thus one never crosses the "disorder line" $h^2 + \gamma^2 = 1$ in the
 $\gamma-h$-plane. In the terminology of \cite{Its04} one is in the regions (1a) and (2).
 In these regions we have completely determined the reduced density matrix.
 Comparing with Ref.\cite{Its04} one sees that the eigenvalues $\varepsilon_j$
 correspond to the quantities $\lambda_j$ there. Moreover, $\varepsilon$ is precisely 
 the quantity $\pi\tau_0$, as can be seen by evaluating the integrals for $\tau_0$.
 The two approaches therefore give the same density-matrix spectrum, as they 
 should, but the way via corner transfer matrices provides more physical insight.\\

  Given the thermal form of the reduced density matrix, the 
 entanglement entropy $S= -tr(\rho\; ln{\rho})$ can immediately be written down
 \be
  S= \sum_{j} ln{\;[1+exp{(-\varepsilon_j)}]} + \sum_{j} \frac {\varepsilon_j}
        {exp{(\varepsilon_j)}+1}
      = lnZ + U
 \label{eqn:S1}
 \ee
 This expression has been evaluated numerically and analysed near criticality
 (where $\varepsilon$ goes to zero) in \cite{Calabrese04}. However, one can
 also obtain a closed form by introducing the elliptic nome
 \be
 q = exp{(-\varepsilon)} = exp{(-\pi\;I(k')/I(k))}
  \label{eqn:q}
 \ee
 and invoking elliptic function identities. In the case $h>1$,
  one finds the following formula for $Z$ from equ.(16.37.3) in \cite{Abramowitz}
 \be 
   \prod^{\infty}_{j=0} (1+q^{2j+1}) = (\frac {16 q} {k^2 k'2})^{1/24}
  \label{eqn:prod}
 \ee
 while for the sum in the "internal energy" $U$, equation (16.23.11) in \cite{Abramowitz} leads
 to
 \be
  \sum^{\infty}_{j=0} (2j+1) \frac {q^{2j+1}} {1+q^{2j+1}} = \frac {1} {24}  [1- (1-2k^2)
                  (2I(k)/\pi)^2]
  \label{eqn:sum}
 \ee
  Similar formulae hold for the case $h<1$. In this way one arrives at the results
  \be
   S=   \frac {1} {12} \left [\;ln{ (\frac {k^2} {16 k'})} + (1-\frac {k^2} {2})
         \frac {4 I(k) I(k')} {\pi} \right ] + ln\;2 ,\;\;\; h<1
    \label{eqn:S2a}
  \ee
  \be
   S=  \frac {1} {24} \left [\;ln{ (\frac {16} {(k^2 k'^2)}} + (k^2-k'^2)
         \frac {4 I(k) I(k')} {\pi} \right ] ,\;\;\; h > 1
   \label{eqn:S2b}
  \ee
 where the $ln\;2$ in (\ref{eqn:S2a}) arises from the eigenvalue $\varepsilon_0 = 0$. 
 Near criticality ($h \rightarrow 1, k \rightarrow 1$) these formulae give the logarithmic
 divergence
 \be
   S \approx \frac {1} {12} \;ln{(\frac {1} {1-k})}
  \label{eqn:S3}
  \ee
 found in \cite{Calabrese04} where the prefactor corresponds to the value $c = 1/2$
 of the central charge.
  For $k \rightarrow 0$, the brackets vanish in both cases.
 If $h \leq 1$, this happens as one approaches the disorder line. Then $|0>$ becomes
 the superposition of two product states \cite{PeschelEmery81} and $S$ goes to
 $ln{\;2}$. The corresponding divergence of the $\varepsilon_j$ for $j \neq 0$, 
 which leads to a collapse of the spectrum of $\rho$, has been seen
 in numerical calculations for small systems \cite{Chung/Peschel01}. A particular
 case is $\gamma = 1$, where one approaches for $h \rightarrow 0$ a superposition
 of the two states with all spins in the positive resp. negative $x-$direction.
 For $h > 1$, the parameter $k$ vanishes as $h \rightarrow \infty$. Then $|0>$ becomes 
 the simple product state $|+++.....>$ and the entropy vanishes.\\
 
 All these results refer to a chain with open ends which is divided into two parts.
 These two parts have one point of contact. If one cuts a segment out of a long 
 chain or a ring, which is the situation studied in \cite{Its04}, one creates $\it{two}$ 
 such interfaces. Since the density matrix eigenfunctions for the
 low $\varepsilon_j$ are localized near an interface, one obtains a double degeneracy
 of the $\varepsilon_j$ with $ j \neq 0$ if the segment is longer than the correlation 
 length. This degeneracy was also noted in \cite{Its04} and 
 gives a factor of two for the brackets in (\ref{eqn:S2a}) and (\ref{eqn:S2b}), while the 
 logarithm is unchanged.\\
 
 We have not discussed the region $\gamma^2+h^2 <1$ inside the disorder line (the
 region (1b) in \cite{Its04}). The CTM's for this case have
 not been studied so far. However, on the line $h = 0$ one can relate the Hamiltonian
 of the XY chain to the transfer matrix of a doubled square lattice, i. e. an eight-vertex 
 model at the decoupling point \cite{Suther70,Baxter82}. For this case the CTM's are 
 known again \cite{Baxter82}. One then obtains $\varepsilon_j = j\;\varepsilon$ and 
 the value for $k$ is $k = (1-\gamma)/(1+\gamma)$. This corresponds again to the 
 result for $\pi \tau_0$ in \cite{Its04}. Since now even and odd multiples of 
 $\varepsilon$ enter, the entropy for a chain cut in half becomes the 
 sum of the two expressions (\ref{eqn:S2a}) and (\ref{eqn:S2b}). The divergence of $S$
 in the isotropic limit $\gamma \rightarrow 0, k \rightarrow 1$ is therefore also
 stronger than (\ref{eqn:S3}) by a factor of two, in accordance with the value
 $c = 1$ for this critical point.\\
 
 $\it{Acknowledgement.}$ The author thanks K. D. Schotte for providing the two figures
 and B. Davies for discussions.

\pagebreak 

\begin{figure}
\centerline{\psfig{file=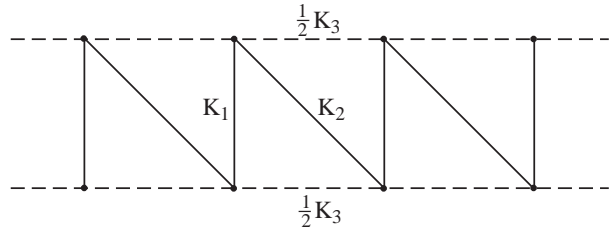,width=8cm,angle=0}}
\vspace{5mm}
\caption{Portion of a triangular lattice forming the row transfer
   matrix T}
\label{fig1.eps}
\end{figure}

\begin{figure}
\centerline{\psfig{file=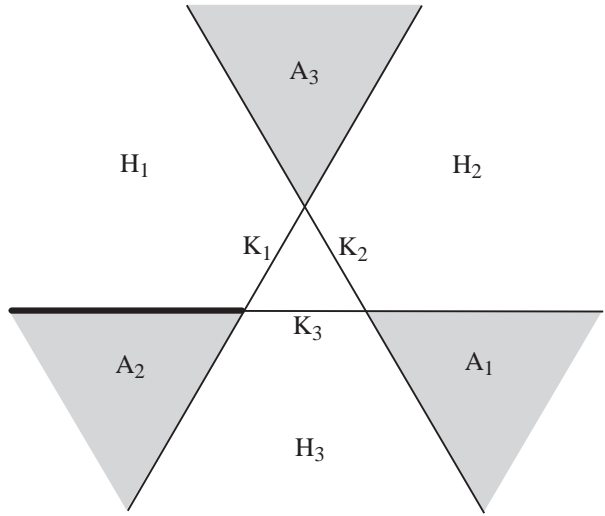,width=8cm,angle=0}}
\vspace{5mm}
\caption{Construction of the reduced density matrix $\hat{\rho}$ from six triangular 
             corner transfer matrices.  Along the heavy line the lattice is cut open.
              The notation follows Ref. [13].}
\label{fig2.eps}
\end{figure}

\end{document}